\documentclass[aip,amsmath,amssymb, reprint]{revtex4-1}
\usepackage{physics}
\usepackage{amsmath}
\usepackage{mathptmx,bm}
\usepackage{graphicx}
\usepackage{dcolumn}
\usepackage{tabularx}
\usepackage{epsf}
\usepackage{color}
\usepackage{hyperref}
\hypersetup{breaklinks,colorlinks,linkcolor=blue,citecolor=blue,urlcolor=blue}

\usepackage{epstopdf}%
\usepackage{verbatim}
\usepackage[english]{babel}
\newcommand{\bea}{\begin{eqnarray}}
	\newcommand{\eea}{\end{eqnarray}}
\usepackage{amsfonts}
\usepackage{graphicx}
\usepackage{dcolumn}
\usepackage{bm}
\usepackage[utf8]{inputenc}
\usepackage[T1]{fontenc}
\usepackage{mathptmx}
\usepackage{etoolbox}
\usepackage{multirow}

\makeatletter
\def\@email#1#2{%
	\endgroup
	\patchcmd{\titleblock@produce}
	{\frontmatter@RRAPformat}
	{\frontmatter@RRAPformat{\produce@RRAP{*#1\href{mailto:#2}{#2}}}\frontmatter@RRAPformat}
	{}{}
}%
\makeatother

\begin{document}

\title{Tuning molecular thermal conductance through endgroup modification and halogen substitution}

\author{Jonathan J. Wang}
\affiliation{Chemical Physics Theory Group, Department of Chemistry, University of Toronto, 80 Saint George St., Toronto, Ontario M5S 3H6, Canada}

\author{Dvira Segal}
\affiliation{Chemical Physics Theory Group, Department of Chemistry, University of Toronto, 80 Saint George St., Toronto, Ontario M5S 3H6, Canada}

\affiliation{Department of Physics, University of Toronto, 60 Saint George St., Toronto, Ontario M5S 1A7, Canada}
\email{dvira.segal@utoronto.ca}

\date{\today}

\begin{abstract}
We demonstrate tuning of the phononic thermal conductance in single molecules with carbon-chain backbones through modifications of terminal groups and halogen substitution of hydrogen atoms. Our simulations focus on intrinsic molecular properties, and we employ a workflow based on {\it ab initio} molecular dynamics, enabling the training and development of machine-learned interatomic potentials. These potentials are subsequently used in classical nonequilibrium molecular dynamics simulations to extract thermal conductance coefficients. Replacing terminal methyl groups with amine, sulfur, or halogen substituents leads to pronounced changes in thermal conductance: bromine-terminated chains exhibit the lowest conductance, whereas amine and methyl-terminated chains show the highest. Additionally, single-atom substitution of hydrogen by fluorine or other halogens along the alkane backbone significantly reduces thermal transport. Finally, our simulations of the length dependence of thermal conductance in alkane chains containing 3–12 carbon atoms reveal its saturation beyond eight carbon atoms. 
Together, our findings show that simple chemical modifications offer a versatile route to controlling phononic heat flow in single molecules.
\end{abstract}

\maketitle
\section{Introduction}



Over the past several decades, both experimental and computational advances in nanoscale thermal transport have enabled systematic manipulation of heat flow, including its controlled enhancement or suppression \cite{RevLu08,Dhar,Pop10,Baowen12,Luo13,Rev14,Leitner15,RevA,Rubtsov19,Yoon20,HuRev21,BaowenR21,Baowen22,RevG,Rev23,Rev25}. Measurements of setups relevant for device applications, such as self-assembled monolayers of organic molecules bridging metal electrodes in steady state \cite{Wang06,Cahill12,GotsmannExp14,Shub15,Shub17}, or in solution using transient-IR pump-probe methods
\cite{Rubtsov19,Rubtsov21,Rubtsov24}
are well established. More recently, advances in instrumentation have allowed the phononic thermal conductance of {\it single-molecule} junctions to be quantified \cite{CuiExp19,GotsmannExp19,GotsmannExp23,CuiExp25}. These experiments remain challenging: whereas electronic conductance of single molecules can vary over many orders of magnitude, the thermal conductance of molecular junctions was found to be typically limited to 5 to 50 pW/K, requiring strict temperature control and careful suppression of fluctuations. 


There is a long history of theoretical and computational studies of phononic thermal conduction in molecules that aim to provide a theoretical basis, mechanistic understanding, and predictions to motivate and explain experiments \cite{RevLu08,Dhar,Baowen12,Luo13,Rev14,Leitner15,RevA,Rev25}. Experiments probe the overall thermal conductance, which arises from the combined contributions of interacting carriers: electrons, phonons, and photons. To isolate the phononic contribution, experiments (specifically on single molecule junction) have focused on molecular systems with poor electronic thermal conductance, such as alkane chains \cite{CuiExp19,GotsmannExp19}. Given the computational challenges of simulating quantum thermal transport from first principles while accounting for both electrons and phonons, a variety of approaches have been developed over the years.
These methods include low-dimensional modeling of single-molecule junctions with a phenomenological treatment of anharmonic effects  \cite{Dhar,Dvira2003,SegalQME,SegalSB,Dhar06,Malay11, Tulkki13,Tulkki15,Roya19,Naim}; fully atomistic calculations based on electronic structure methods, typically limited to harmonic approximation and coherent scattering using the nonequilibrium Green's function (NEGF) formalism \cite{Markussen13,Pauly16,ChristmasTree,Teflon,Pauly17,Pauly18,Pauly23,Hatef25,Heavyatom,Fabian25}, which can also provide electronic contribution to thermal conductance; and classical molecular dynamics (MD) simulations, most often in nonequilibrium settings (NEMD), taking into account anharmonic effects with empirical potentials \cite{Pawel11,NitzanD20,Nitzan20,Lu2021,Nitzan22,Benzene}. 
%
Specifically, in recent works, we used classical MD simulations with empirical force fields to determine the phononic thermal conductance of carbon based single-molecule junctions \cite{JW-HeatMD,JW-HeatFluc,JW-Fullerene}. We focused on the well-studied gold–alkanedithiol–gold junction \cite{JW-HeatMD,JW-HeatFluc}, and our calculated thermal conductances agreed with both previous theoretical results and experimental measurements. 

A central question in nanoscale thermal transport concerns structure–function relationships: which specific molecular or junction components can enhance or suppress heat flow. In principle, control over molecular thermal transport can be achieved through modifications of the backbone, terminal groups, side groups, or metal contacts.
However, classical MD simulations using empirical potentials rely on parameterizations that often lack consistency and accuracy. To achieve both flexibility and reliability across a wide range of molecular systems, we adopt machine-learned interatomic potentials (MLPs) in this work \cite{ANI,DeeP,MLP19,MLthermal20,MLRev21,MLrev1}. Recent developments in MD methodology have leveraged machine learning to combine the accuracy of quantum mechanical calculations with the efficiency of classical MD.  Quantum calculations typically involve density functional theory (DFT) to determine potential energies and forces arising from electronic interactions, which are then used in {\it ab initio} molecular dynamics (AIMD) to propagate nuclear motion. Because AIMD is computationally expensive, simulations are usually limited to picosecond timescales, whereas classical MD can reach nanoseconds or longer. MLPs aim to combine the strengths of both approaches: the accuracy and consistency of AIMD with the extended timescales accessible in classical MD \cite{ANI,DeeP,MLP19,MLthermal20,MLRev21,MLrev1}.

Although MLPs have found many applications in chemistry and material science, for example, for high-throughput screening and property prediction \cite{MLrev1}, only a few studies use MLPs to study thermal transport of single molecules: 
One such example is the study in Ref.~\citenum{Lu21}, which examined the heat-transport properties of alkane junctions with gold or graphene electrodes using AIMD-trained machine-learning potentials.


The objective of this work is to investigate the thermal conductance of single-molecule systems towards understanding of the structure–function relationships that govern nanoscale heat transport. 
Our workflow combines {\it ab initio} molecular dynamics to generate short trajectories for training machine-learned interatomic potentials, which are then used in nonequilibrium molecular dynamics simulations to compute phononic thermal conductance. We focus on molecules with carbon backbones. Experimentally, thermal conductance is measured in molecular junctions, where the molecule is connected to metal electrodes via endgroups such as thiol or amine groups. However, as a first step in this study, we simulate only the molecule itself, end-to-end, thus omitting metal–molecule contact resistance and focusing solely on the intrinsic thermal properties of the molecule.
We focus on two key aspects of thermal transport in alkane chains: the influence of terminal groups and the effect of halogen substitutions along the alkane backbone. In addition, we examine the length dependence of the intrinsic (molecular) thermal conductance of alkane chains. Experimental studies of alkanes as molecular junctions have reported both length-independent \cite{CuiExp19} and length-dependent trends \cite{GotsmannExp14}, motivating our computational investigation.

The plan of this paper is as follows. In Sec. \ref{Method-sec}
we describe the simulation procedure. In Sec. \ref{End-sec} we study the effect of different endgroups on thermal conductance. The impact of halogen substitution is explored in Sec. \ref{Halo-sec}. We examine the length dependence of conductance in alkane chains in Sec. \ref{Length-sec} and conclude in Sec. \ref{Sum-sec}.
\begin{figure}
    \centering
\includegraphics[width=\linewidth]{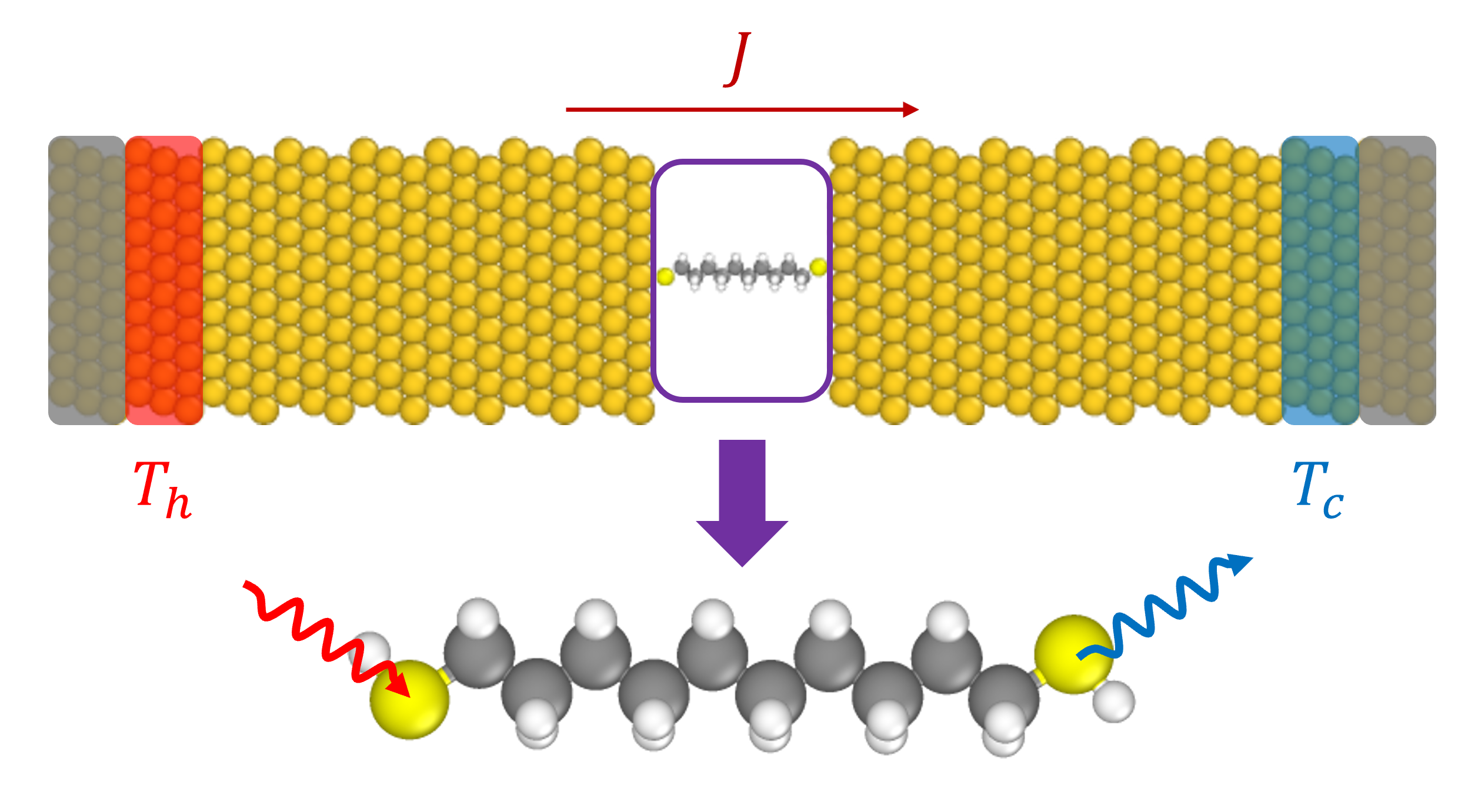}
\caption{The gold–alkanedithiol–gold single-molecule junction represents a typical setup for experiments and MD simulations such as in  Refs. \citenum{JW-HeatMD, JW-HeatFluc}. In the schematic, the gray regions at both ends indicate fixed atoms, while the red and blue regions correspond to thermostatted atoms maintained at target temperatures, $T_h$ and $T_c$, respectively. Heat current $J$ develops through the system, allowing the thermal conductance to be determined. The purple box highlights the molecule, which is the primary focus of this study. In this work, we investigate heat flow considering the  molecular structure alone, with the thermostats applied directly to the molecular terminal atoms, indicated by red and blue arrows.}
    \label{Fig1}
\end{figure}


\section{Setup and Simulation Workflow}
\label{Method-sec}


To probe thermal conductance at the single-molecule level, experiments have employed molecular junctions, with the connecting electrodes playing a critical role in the measured value. In Refs. \citenum{JW-HeatMD, JW-HeatFluc}, we used classical MD simulations to study the thermal conductance in a corresponding gold–alkanedithiol–gold junction, as illustrated in Fig. \ref{Fig1}. Thermal conductance computed using different flavors of NEMD methods agreed well with previous computational and experimental results for this junction. However, extending this approach to a wider variety of molecules would require seeking force field parameters, if empirical interatomic potentials were continued to be used. To address this issue, we adopt here machine-learned potentials trained on {\it ab initio} molecular dynamics data. This allows us to maintain a consistent methodology while exploring diverse molecular systems and capturing thermal transport at classical MD timescales.

Ab-initio calculations for molecular junctions are computationally demanding due to the need for separate optimizations for system size, basis sets, convergence thresholds, and other parameters between metal and molecule. Moreover, AIMD simulations require self-consistent field calculations at every MD timestep, further increasing the computational cost. Given these challenges in generating sufficient AIMD data for complete junctions, in this study, we do not include the metal contact in simulations but focus on the molecular conductance rather than the overall junction's behavior, see Fig. \ref{Fig1}. 

In brief, the procedure for calculating the thermal conductance of molecules involves generating machine-learned potentials and using them in nonequilibrium MD simulations. This procedure consists of three main components: (1) AIMD simulations performed with the SIESTA program \cite{Siesta1, Siesta2} to generate training data for the ML models; (2) training of MLPs using the DeePMD-kit \cite{DPMD1, DPMD2}, where multiple models are trained for each molecule to ensure reliability during MD simulations; and (3) performing two-step production MD simulations in LAMMPS \cite{LAMMPS} with the trained MLPs used to simulate the thermal conductance. Visualization of atomic coordinates and MD trajectories are done by OVITO\cite{OVITO}. Specific details of each step are described next.

{\bf (1) AIMD simulations:} We perform simulations in SIESTA \cite{Siesta1,Siesta2} employing the Nosé–Hoover thermostat to control the system temperature, sampling the canonical ensemble (NVT). Training data for ML models are collected from simulations at multiple temperatures: 50 K, 100 K, 150 K, 200 K, and 300 K. Each simulation produced 2 ps trajectories with a timestep of 0.5 fs, recording atomic coordinates, energies, and forces from self-consistent field calculations at every step. These outputs serve as the `frames' for MLP training. To reduce correlations between frames, the data were downsampled by selecting every 20th frame, corresponding to 10 fs intervals in simulation time.

{\bf (2) MLP training:} For each molecule, the data set consists of 1000 frames collected from the NVT AIMD simulations at different temperatures. Of these, 800 frames are used for training and 200 for testing. Using the DeePMD-kit, deep neural network (NN) models are constructed employing the two-body embedding DeepPot-SE descriptor to fit the potential energy \cite{DPMD1,DPMD2}. For each molecule, a baseline of four independent models is trained, each with different initial parameters and variations in frame selection to ensure robustness. Additional details on the ML training procedure are provided in Appendix A. 
In Appendix B, we display simulated vibrational spectra of examined molecules based on their MLPs, where we identify several known modes.

{\bf (3) NEMD simulations:} The 
Large-scale Atomic/Molecular Massively Parallel Simulator (LAMMPS) \cite{LAMMPS} is integrated with the DeePMD-kit to directly use trained MLPs in MD simulations. Thermal conductance is determined through two MD runs: First, the system is equilibrated under the NVT ensemble (Nosé–Hoover thermostat) at the target average temperature $\bar{T}$ for 1 ns with a 1 fs timestep. In the subsequent NEMD production run, Langevin thermostats are applied to the two ends of the molecule (Fig. \ref{Fig1}) during a 2 ns NVE simulation at 1 fs timestep. Unless otherwise specified, the NVT equilibration targets $\bar{T} = (T_h + T_c)/2 = 100$ K, and the NVE run applies $T_h = 125$ K and $T_c = 75$ K.

The energy exchange $\Delta E_{h,c}$ between the thermostatted ends and the Langevin baths is recorded over a time interval $t$ to calculate the net heat current, $J = |\Delta E_{h,c}|/t$, while verifying steady-state flow. The per-atom kinetic energies were also monitored to construct a time-averaged molecular temperature profile. In practice, the temperature averaged over time of the thermostatted atoms, $T_L$ and $T_R$, often deviate from the target, $T_h$ and $T_c$. We therefore define the actual temperature difference as $\Delta T = T_L - T_R$ for each simulation. The thermal conductance is calculated as $G_{\rm th} = J / \Delta T$. Error estimates are obtained by repeating simulations with different trained MLP models and setting random initial velocities from the canonical distribution to initialize NEMD simulations. 

We address a common challenge in applying MLPs: insufficient sampling of the molecule’s configurational space can lead to stability issues during classical MD simulations \cite{DPMD1,DPMD2}. In our workflow, the MLP is trained to predict forces and energies for structures sampled from AIMD trajectories. Through preliminary trials, we developed a procedure 
which was applied consistently across all examined molecules, see Appendix A. Importantly, all results presented here include repeated simulations for error estimation, and are considered ``stable". Instabilities were identified in early trials as cases in which the molecular structure broke apart during MD simulations in LAMMPS, which we excluded in the final data set.

Molecular breakdown during the MD stage arises from insufficient sampling, i.e., an incomplete representation of the potential energy surface, particularly for configurations involving significant short-range interactions. For example, the base structure of our examined molecules is a linear chain. In MD trajectories, we occasionally observed configurations in which the chains bend sharply, bringing one end close to the middle. Due to the much shorter timescales accessible in AIMD, such configurations were rarely sampled then, and thus the MLP did not learn to handle them. When encountered during MD, incorrect estimation of repulsive forces could cause an abrupt increase in kinetic energy, leading to structural breakdown. A straightforward way to mitigate this would be to extend AIMD simulations to capture more configurations. However, to keep the study within scope, we acknowledge this limitation and present only results that are confirmed to be stable within a few nanoseconds of trajectory.

\section{Effect of molecular endgroups on thermal transport}
\label{End-sec}

\begin{figure}
    \centering
    \includegraphics[width=\linewidth]{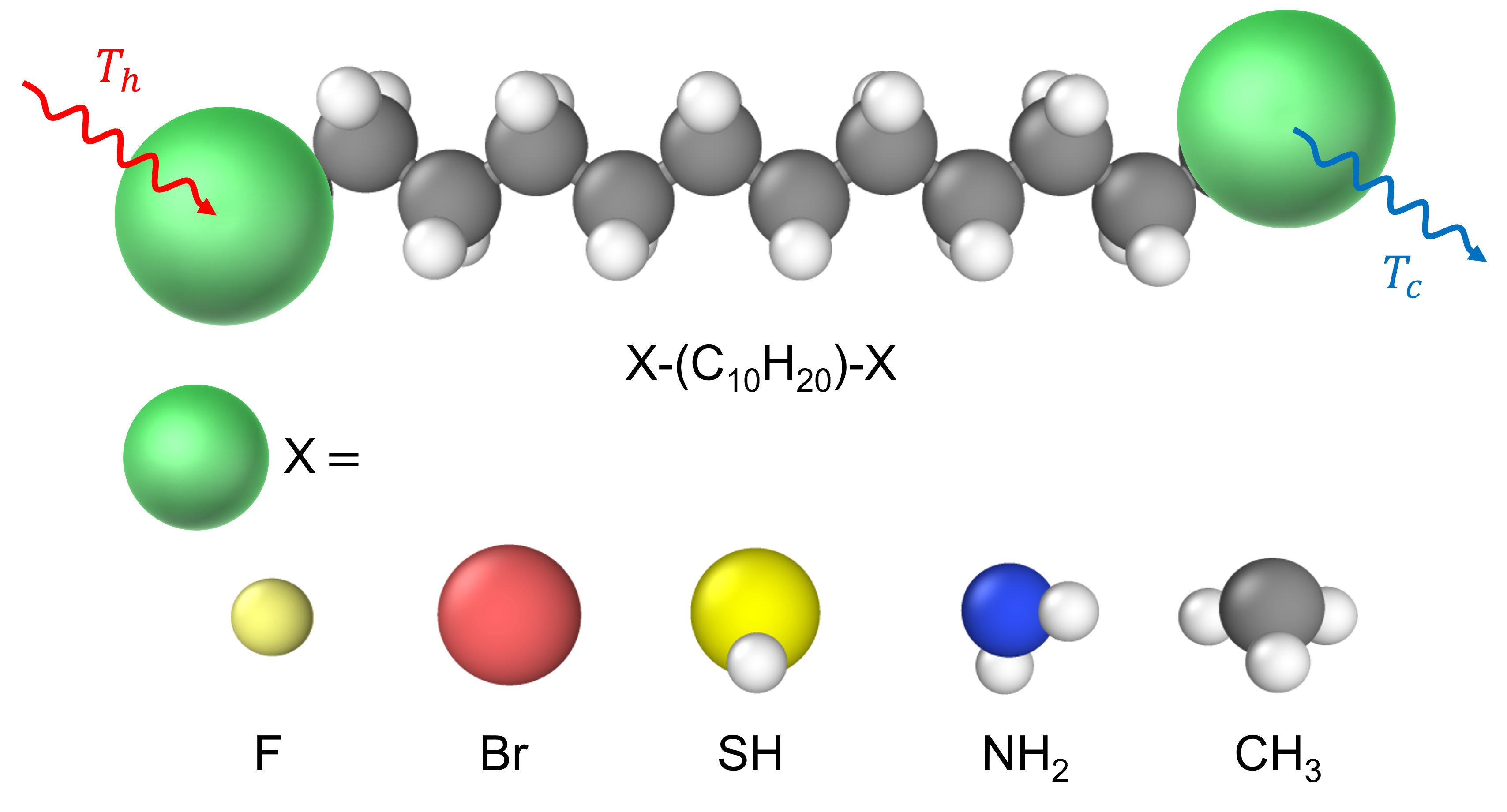}
    \caption{Schematic of a decane ( XC$_{10}$H$_{20}$X) system used to examine the effect of different molecular endgroups (X). In the NEMD production run, Langevin thermostats are applied to the central atom of each endgroup, maintaining a hot temperature $T_h$ and a cold temperature $T_c$ at opposite ends of the molecule. Endgroups are depicted to relative scale.}
    \label{Fig2}
\end{figure}

\begin{figure*}
    \centering
    \includegraphics[width=\linewidth]{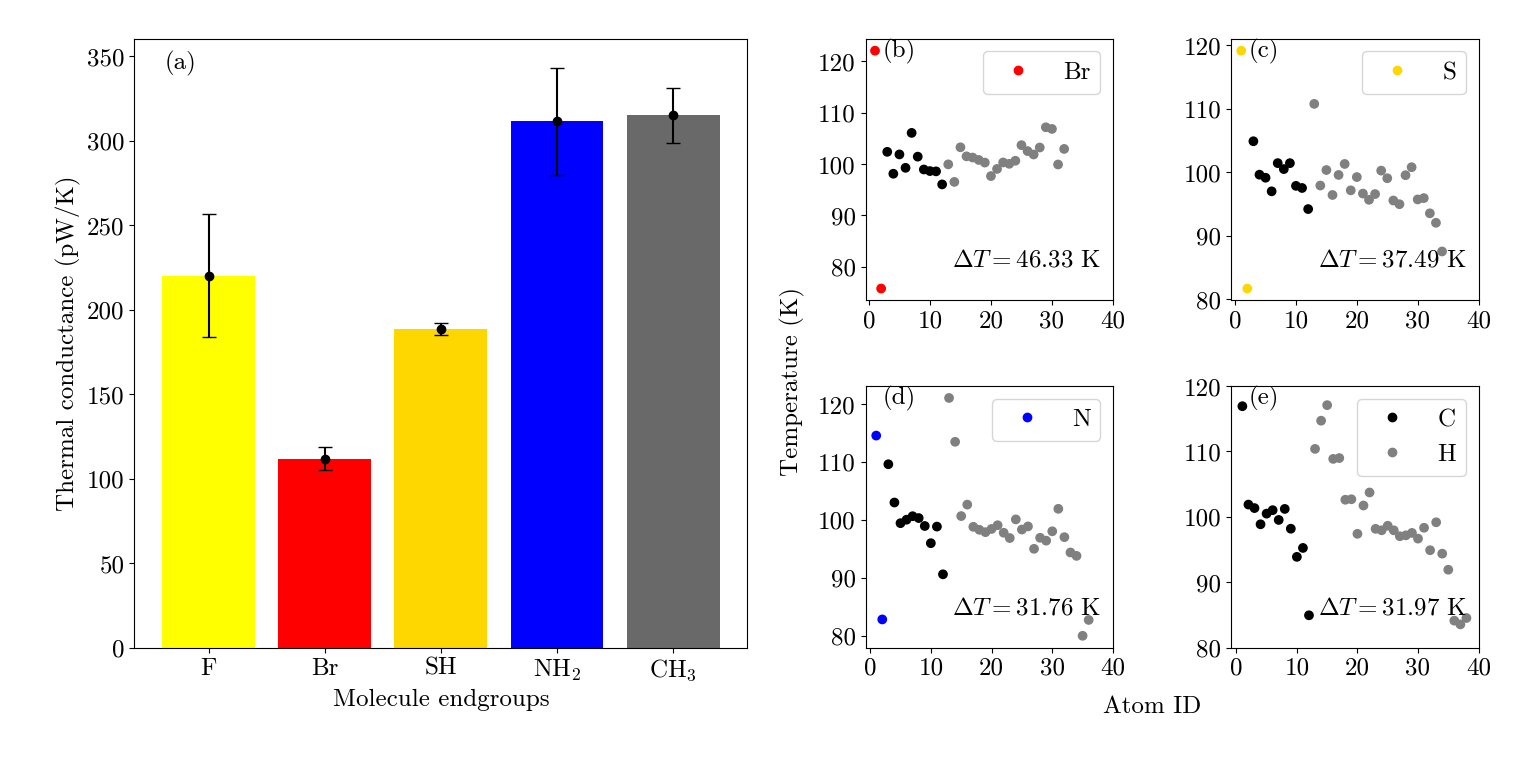}
    \caption{{\bf (a)} Thermal conductance results of C$_{10}$H$_{20}$ alkane chains with different thermostatted contact groups at the ends. {\bf (b)-(e)} Select temperature profiles from individual production runs for endgroups of (b) Br, (c) SH, (d) NH$_2$, and (e) CH$_3$. Temperature difference $\Delta T$ are calculated as the difference between temperatures of terminal atoms.
}
    \label{Fig3}
\end{figure*}

It is well established that, in molecular junctions, the metal–molecule contact plays a critical role in determining transport properties: Depending on the strength and nature of the interaction, a contact resistance arises at the electrode–molecule interface, which can be comparable to or even exceed the intrinsic molecular resistance. This effect has been extensively investigated in both electronic \cite{electronG} and phononic thermal transport \cite{Baowen22}. In particular, experiments on self-assembled monolayers have examined the influence of end groups, showing that chemisorbed contacts yield conductance higher than that of physisorbed contacts \cite{Cahill12}.
Thermal boundary conductance was also demonstrated computationally, e.g., in alkane chains placed between various metal substrates  \cite{Nitzan22}, molecules consisting of two moieties, anthracene and azulene, with a polyethylene glycol oligomer bridge \cite{Leitner19} and in biomolecular systems \cite{LeitnerB}.

In the present work, we investigate the influence of different molecular endgroups attached to an alkane chain of a fixed length. Our focus is specifically on the intramolecular contact resistance that arises between the endgroup and the alkane backbone. The external resistance associated with metal–endgroup coupling is not considered in our analysis. 

The structure of the system is illustrated in Fig. \ref{Fig2}, where five different endgroups are studied. The choice of endgroups is motivated by commonly studied molecular junctions, where they serve as contacts with electrodes, e.g., gold–sulfur or graphene–carbon interfaces. All molecules considered here are neutral; for certain endgroups (SH, NH$_2$, CH$_3$), covalent coupling to electrodes would occur in a junction setup. In our simulations, Langevin thermostats are applied to the end atoms during the production run, specifically to the central atom of the endgroup if it contains hydrogens. The terminal atoms continuously exchange energy with the thermostats, which is recorded cumulatively to determine the heat current at nonequilibrium steady state. Using the time-averaged atomic kinetic energies to calculate $\Delta T = T_L - T_R$, we compute the thermal conductance $G_{\rm th}$ for each molecule.

The results of the thermal conductance are compiled in Fig. \ref{Fig3}(a). The primary observation is that molecular thermal transport in this system is strongly influenced by the interaction between the inner alkane chain and the contacting endgroups. In these molecules, vibrational heat transport occurs predominantly through the C–C bonds of the backbone, while the high-frequency C–H modes are localized and contribute little to heat transport. 
Previous computational studies, such as Refs. \cite{Shub15, JW-HeatMD, JW-HeatFluc}, even exclude explicit hydrogen atoms, treating CH$_2$ groups as single interaction sites.

Fig. \ref{Fig3}(a) shows trends in thermal conductance that reflect the combined effects of mass, atomic radii, and bonding characteristics when attaching the C$_{10}$H$_{20}$ backbone to various endgroups. For example, CH$_3$ endgroups correspond to a fully extended alkane chain (dodecane), where matching backbone structure and C–C vibrational frequencies enable efficient, unhindered transport. NH$_2$ endgroups exhibit similar conductance, likely due to comparable size, bonding characteristics, and vibrational frequencies of the C–N bond. In contrast, the thermal conductance decreases significantly for F, Br, and SH endgroups compared to the plain alkane and the amine termination.

For fluorine, the strong and short C–F bond, resulting from the high electronegativity of F, leads to a vibrational frequency mismatch with the C–C backbone, impeding energy transfer. The weaker and longer C–Br and C–S bonds similarly produce vibrational mismatches that hinder thermal transport. The particularly low conductance for the heavier endgroups is also consistent with their increased mass and size relative to the backbone atoms (C, N, and F), which further suppresses the flow of vibrational energy.

Fig. \ref{Fig3}(b)–(e) shows the time-averaged temperature profiles of the examined molecules (excluding the F endgroup), plotted along the molecular chain from left to right. Atoms are grouped by element type. Therefore, in Fig. \ref{Fig3}(b)–(d), atom IDs 1 and 2 correspond to the thermostatted core atoms of the endgroups, whereas in the pure alkane chain (e), atom IDs 1 and 12 represent the thermostatted ends. The temperature difference between these atoms provides $\Delta T$, shown in the insets, which is used to calculate the thermal conductance in Fig. \ref{Fig3}(a).

The profiles are ordered according to increasing thermal conductance, revealing trends consistent with the simulated behavior. For example, in Fig. \ref{Fig3}(b), the temperatures of the two Br atoms are closest to the thermostat target values (125 K and 75 K), producing the largest $\Delta T$. In contrast, $\Delta T$ decreases to $\sim$32 K for NH$_2$ and CH$_3$ endgroups. These results indicate that CH$_3$ and NH$_2$ endgroups thermalize effectively with the alkane backbone, reflecting efficient vibrational transport, whereas Br and SH endgroups exhibit poor thermal coupling with the backbone, causing their temperatures to remain closer to the thermostat targets.

Direct comparison of the thermal conductance values obtained here with previous studies is not straightforward, as our simulations consider isolated molecules rather than full junctions with explicit electrodes. As shown in Fig.~\ref{Fig3}, the intrinsic molecular $G_{\rm th}$ values computed in this work are roughly an order of magnitude higher than those typically calculated for single-molecule junctions \cite{JW-HeatMD}, where in the latter, the conductance is reduced by contact resistance. 
Calculations of phonon thermal conductance in {\it junctions} with the same molecule type, specifically S or NH$_2$ endgroups in contact with gold, displayed the opposite trend reported here, where S contacts resulted in slightly higher conductance, probably due to its stronger attachment to gold, reflecting the importance of the metal-contact interaction when studying molecular junctions \cite{Pauly16}.
Comparable conductances to what we obtained here for molecule-only alkane systems have been reported in Ref. \cite{ChristmasTree} based on NEGF calculations. Consequently, our results are most useful for comparative analysis across different molecules, providing insight into how chemical and structural variations influence thermal transport.


\section{Effect of halogen substitution}
\label{Halo-sec}

\begin{figure*}
    \centering
    \includegraphics[width=\linewidth]{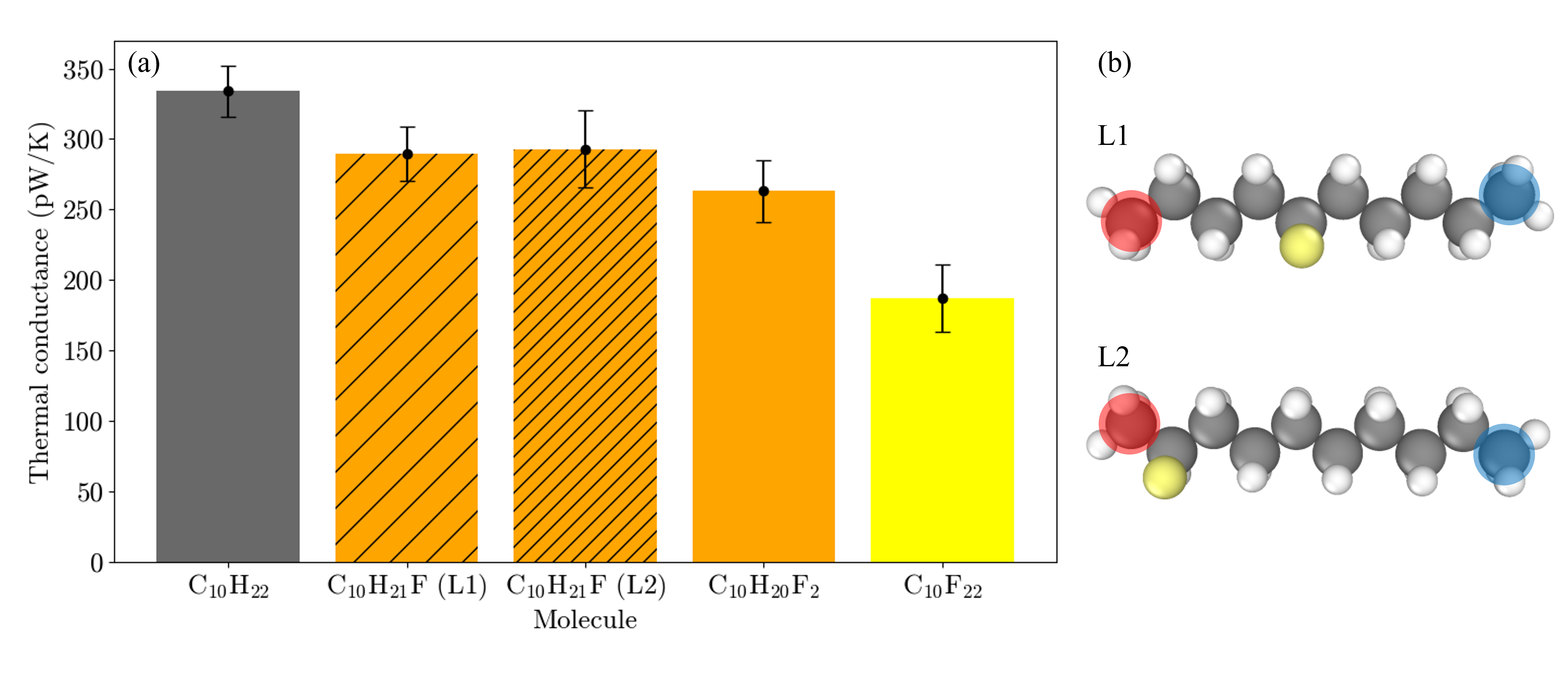}
    \caption{{\bf (a)} Thermal conductance of carbon-based molecules with F substitution. (Left-to right) As a reference point we begin from pure decane that has H atoms (gray). A single fluorine substitution on the alkane backbone is probed at two locations (L1 and L2, hashed orange). Other structures are a chain with two F atoms substituting H at the same C atom (orange), and a fully fluorinated perfluorodecane (yellow). {\bf (b)} The structures of C$_{10}$H$_{21}$F differing by F substitution location L1 and L2. The F atom is colored in yellow; red/blue colorings indicate thermostatted C atoms.}
    \label{Fig4}
\end{figure*}

\begin{figure}
    \centering
    \includegraphics[width=\linewidth]{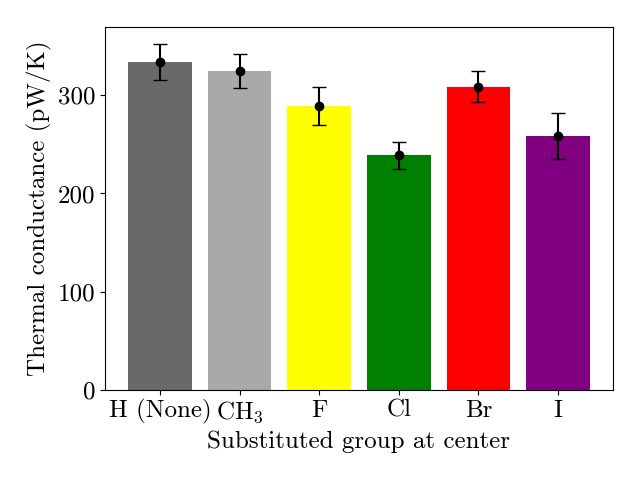}
    \caption{Thermal conductance of alkane chains with a single substitution 
    on the C$_{10}$ backbone at the center of the chain (position L1 from Fig. \ref{Fig4}(b)).  Molecules are C$_{10}$H$_{21}$X where X stands for H (no substitution); CH$_3$ group; a halogen atom. }
    \label{Fig5}
\end{figure}

We next examine thermal transport in systems that share the same alkane backbone but have their hydrogen atoms substituted by a different element. As discussed in the previous section, the C–H motion within each methylene unit are highly localized and therefore contribute minimally to energy transport along the C–C chain. This arises from the low mass of hydrogen, which leads to high-frequency vibrational modes that weakly couple to backbone vibrations. It is unclear to what extent replacing H with a heavier element could modify the thermal conductance. One might expect that increasing the mass of the side group could shift its C–X (with X being the substituent atom) stretching modes to frequencies more comparable to the C–C stretching modes of the backbone. This near-resonance situation could enhance scattering between backbone and side-group vibrations, thereby suppressing heat transport along the chain.
To explore this effect, we first consider a single fluorine substitution. After that, we study alkanes with two fluorine substitutions, then perfluoroalkanes, which were chosen because they are well studied, chemically stable, and represent the limit of complete substitution of hydrogen atoms in the alkane framework.

Figure \ref{Fig4}(a) shows the thermal conductance for various degrees of fluorine substitution. In all MD simulations for this section, the C$_{10}$ backbone is kept fixed, and thermostats are attached to the two terminal carbons. We first consider single-atom fluorine substitution at two different backbone sites (hashed orange bars). The specific locations, labeled L1 and L2, are illustrated in Fig. \ref{Fig4}(b). In L1, the substituted carbon is near the center of the chain, whereas in L2 the substitution occurs near the end, adjacent to the thermostatted boundary.
For a single F substitution, the molecular thermal conductance
$G_{\rm th}$ decreases by roughly 15\% relative to the pure alkane chain, with no significant difference between the two substitution sites. This indicates that replacing even a single hydrogen with a heavier atom, shifting a C–H vibrational mode to a lower frequency, reduces the molecule’s thermal transport capability.

This observation motivated us to further increase the degree of fluorination. The next case, shown as the flat orange bar, involves two F atoms substituted on the same carbon, effectively replacing a methylene group with a difluorocarbene unit. This substitution was performed at location L1. The resulting $G_{\rm th}$ continues the observed trend, decreasing by an additional approximate 10\% relative to the single-F substitution.
We note however that the C$_{10}$H$_{20}$F$_2$  molecule proved significantly less stable in our MD simulations than the other structures studied. As a result, converged conductance values were obtained from only two trained MLPs, compared to four for the other molecules.

Finally, we report the thermal conductance of the fully fluorinated perfluorodecane, whose $G_{\rm th}$
is roughly half that of decane. Together, the fluorination results indicate that vibrational heat transport is progressively hindered as hydrogen atoms are replaced by fluorine. Instead of efficiently propagating heat along the carbon backbone, as occurs through the pure alkane chain, the vibrational energy is likely scattered by the interactions with the fluorine. 
Fully fluorinated alkane chains were also examined in a junction setup with gold electrodes when using NEGF calculations \cite{Pauly16}, showing a reduction in thermal conductance compared to the regular alkane chains, although the degree of reduction varied depending on the metal-molecule contact. 

Next, we examine how thermal transport is affected when a single hydrogen atom at the L1 site is replaced by different substituent groups, specifically a CH$_3$ group or various halogens. The resulting thermal conductances are shown in Fig. \ref{Fig5}, which also includes the H- and F-substituted cases from Fig. \ref{Fig4}  for comparison.
We first consider the effect of introducing a branching alkane by attaching a CH$_3$ side group 
to the backbone. This substitution produces a slight decrease in
$G_{\rm th}$, which may reflect the presence of an additional vibrational pathway that disrupts heat transport along the main chain.
As one may expect, the reduction induced by the methyl substitution, using atoms already present in the backbone, is smaller than the reduction observed for the halogen substitutions. Notably, however, when comparing the halogens from F to I, no clear monotonic trend emerges with increasing atomic mass or size.
One may hypothesize that increasing the size of the substituent, effectively introducing a larger defect into the primary heat‐conduction path, would lead to a systematic decrease in $G_{\rm th}$. However, the thermal conductance obtained for the Br-substituted molecule shows that mass or atomic size is not the dominant factor governing $G_{\rm th}$ across the halogen series. Indeed, the vibrational frequencies of C-Cl, C-Br, and C-I are similar, contrary to their masses. 


\section{Length dependence in alkane chains}
\label{Length-sec}

\begin{figure}
    \centering
    \includegraphics[width=\linewidth]{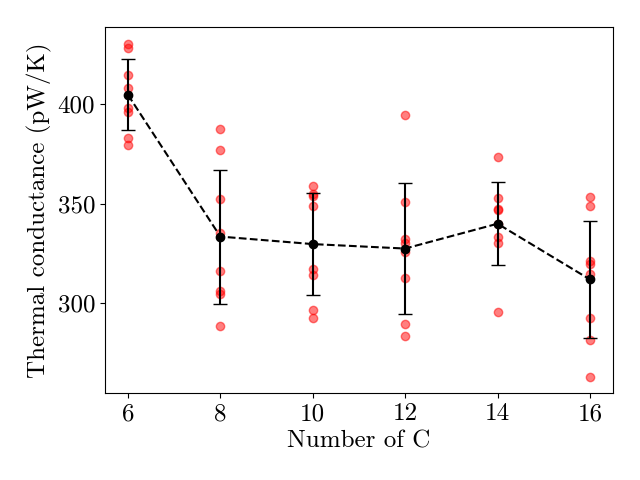}
    \caption{Thermal conductance of alkane chains (C$_{N}$H$_{2N+2}$) as a function of number of carbons ($N$). Red circles for each molecule depicts $G_{\rm th}$ results from eight simulations; black circles display the mean $\langle G_{\rm th}\rangle$ with standard deviation as error bars.}
    \label{Fig6}
\end{figure}

Length-dependent thermal transport in nanoscale systems has been studied computationally in molecular junctions, particularly in alkanes,  \cite{Dvira2003,Pawel11,Pauly16,Roya19,NitzanD20,Nitzan20,Lu2021,JW-HeatMD}. 
Experimentally, contradictory results exist:
Measurements with a scanning thermal microscope with a silicon tip on self-assembled monolayers found the conductance to vary as a function of the alkane chain length, up to a factor of 3 \cite{GotsmannExp14}. The maximum thermal conductance appeared for a chain with four carbon atoms. In contrast, single-molecule experiments on Au–alkanedithiol–Au junctions reported length-independent thermal conductance \cite{CuiExp19}.
Computational results show a similar diversity: empirical harmonic calculations indicate a length dependence \cite{Dvira2003}, whereas {\it ab initio} simulations suggest conductance is largely independent of chain length \cite{CuiExp19}. These discrepancies highlight the critical role of contact thermal resistance in both experiments and simulations, which must be carefully accounted for to meaningfully compare results.


In this section, we simulate the thermal conductance of an alkane chain (C$_N$H$_{2N+2}$) as a function of molecular length. As before, we do not include the metal contact and our calculations provide the intrinsic molecular conductance.
We perform the analysis using our MLP-MD framework, and 
as in Sec. \ref{End-sec}, thermostats are applied only to the terminal carbon atoms.
For each chain length, we collect $G_{\rm th}$ 
values from eight production simulations, each using an independently trained MLP. These individual conductance values are plotted in red in Fig. \ref{Fig6} to illustrate their spread.
The corresponding averages $\langle G_{\rm th}\rangle$ are shown in black. Overall, results reveal only a weak dependence on chain length for $N\geq 8$, consistent with nearly ballistic transport as the chains become longer.

This behavior is consistent with previous molecular-junction studies, which similarly report almost length-independent thermal conductance at long chain lengths, a signature of nearly ballistic heat transport arising from the relatively harmonic nature of C–C bonds in alkanes.
Our results show an enhanced $G_{\rm th}$ at short chains, notably for C$_6$H$_{14}$. Both fully harmonic computational models \cite{Dvira2003} and experimental measurements \cite{GotsmannExp14} have confirmed this elevated short-chain conductance in alkane-based junctions.  

As a side note, in Appendix C we assess the generalizability of our MLPs by applying a model trained on a 12-carbon dodecane to simulate thermal transport in both shorter and longer chains. We find that the dodecane MLP reproduces the qualitative trend of decreasing thermal conductance with increasing chain length, followed by saturation for longer chains. However, the absolute errors in the conductance values are relatively large, indicating that quantitative predictions remain sensitive to the specific training set.

\section{Discussion and Summary}
\label{Sum-sec}

\subsection{Summary of Observations}

In this study, we computationally analyzed thermal transport in alkane-based molecules, focusing on three aspects: (1) the influence of different endgroups; (2) the effect of substituting hydrogen atoms on the alkane backbone with halogens; and (3) the dependence of thermal conductance on chain length.
We quantified the thermal conductance $G_{\rm th}$ through a multistep simulation pipeline: starting with \emph{ab initio} molecular dynamics, followed by training deep neural network models as machine-learning potentials, and finally performing nonequilibrium molecular dynamics simulations of heat transport.

Thermal transport analysis with various endgroups elucidated that the highest thermal conductance occurred for CH$_3$
endgroups, which matched the alkane backbone, while other endgroups lead to progressively lower conductance in the order NH$_2$, F, SH, and Br. 
When substituting hydrogen atoms on the alkane backbone with other substituents including CH$_3$ and halogens, we found that $G_{\rm th}$ decreased compared to the normal alkane chain, possibly due to scattering processes that hinder transport along the backbone.
Finally, in pure alkanes, the thermal conductance is largely independent of chain length for $N\gtrsim8$. 

It is important to emphasize that our results primarily elucidate the relative importance of different factors and underlying mechanisms of molecular thermal transport, rather than absolute conductance values: In realistic devices, absolute thermal conductance will be further affected by the metal–molecule contact. 

\subsection{Insights}
A unifying insight is that phonon transport is controlled by matching of vibrational modes. In more details:

{\bf (1) Endgroup effects: } Thermal conductance is strongly influenced by the type of terminal-thermostatted group attached to the alkane backbone. CH$_3$ and NH$_2$ endgroups yield higher conductance, while heavier or mismatched groups (Br, SH, F) reduce it.
This can be understood by considering the endgroups as interfaces between the molecular backbone and the heat source or sink. Efficient energy transfer requires vibrational mode matching between the backbone and the endgroup to minimize phonon scatterings. The endgroups modify the local vibrational modes, and as such play a central role in controlling how effectively energy is transferred along the molecule.


{\bf (2) Side-group substitution (halogenation):}
Replacing hydrogens with heavier atoms (F, Cl, Br, I) decreases thermal conductance; a single F substitution reduces it by 15\%, while full fluorination roughly halves it. Heavier substituents introduce additional low-frequency vibrational modes that act as phonon scatterers. This demonstrates that chemical modifications along the backbone can be used to tune thermal transport by engineering the local vibrational spectra,  controlling phonon scattering and energy flow along the molecule.


{\bf (3) Machine-learned potentials for thermal transport:} MLPs trained on AIMD data enable modeling of variety of molecules. 
Representing forces accurately is essential to capture nuanced energy-transfer mechanisms such as due to mode mismatch and phonon scatterings. The generalizability of MLPs remains a challenge as well as the capture of trends at high temperatures. Despite these limitations, quantum-informed classical simulations---using trained MLPs---may be able to capture key physics of molecular thermal transport.



\subsection{Future work}
Future work will focus on several directions in molecular heat transport. As a natural extension to the project, we first aim to include explicit metal contacts in AIMD and MLP-based simulations to investigate how the molecule-electrode coupling affects thermal conductance. Second, simulations at elevated temperatures will allow us to probe anharmonic effects and thermal scattering processes that become significant beyond the low-temperature, nearly harmonic regime. More broadly, we plan to study molecules with other backbones such as conjugated chains or branched structures. 

An intriguing objective for future studies is the design of molecular systems that exhibit a transition from ballistic to diffusive heat transport. By systematically tuning, e.g., side-group substitution and chain length, it may be possible to enhance inelastic phonon scattering, thereby exploring the crossover from ballistic, phase-coherent transport to fully diffusive behavior \cite{Dhar,Dhar06, Malay11, Roya19}. 
Together, these studies will provide a more comprehensive picture of structure–function relationships in molecular thermal transport and guide the design of molecules for thermal management applications.


\begin{acknowledgements}
DS acknowledges support from an NSERC Discovery Grant.
The work of JJW was supported by an NSERC Postgraduate Scholarship---Doctoral (PGS-D). 
We acknowledge Longji Cui for discussions that motivated this work.
Computations were performed on the Niagara supercomputer at the SciNet HPC Consortium. SciNet is funded by: the Canada Foundation for Innovation; the Government of Ontario; the Ontario Research Fund---Research Excellence; and the University of Toronto.

\end{acknowledgements}

\vspace{5mm}
\noindent {\bf AUTHOR DECLARATIONS} \\

\noindent {\bf Conflict of Interest}

\noindent The authors have no conflicts to disclose

\vspace{5mm}

\noindent {\bf DATA AVAILABILITY} \\
The data that support the findings of this study are available from the corresponding author upon reasonable request.


\renewcommand{\theequation}{A\arabic{equation}}
\renewcommand{\thesection}{A\arabic{section}}
\renewcommand{\thesubsection}{A\arabic{subsection}}
\setcounter{equation}{0}
\setcounter{section}{0} 
\setcounter{subsection}{0} 

\section*{Appendix A: Machine learning procedure and testing}
\label{app:1}

\begin{figure*}
    \centering
    \includegraphics[width=\linewidth]{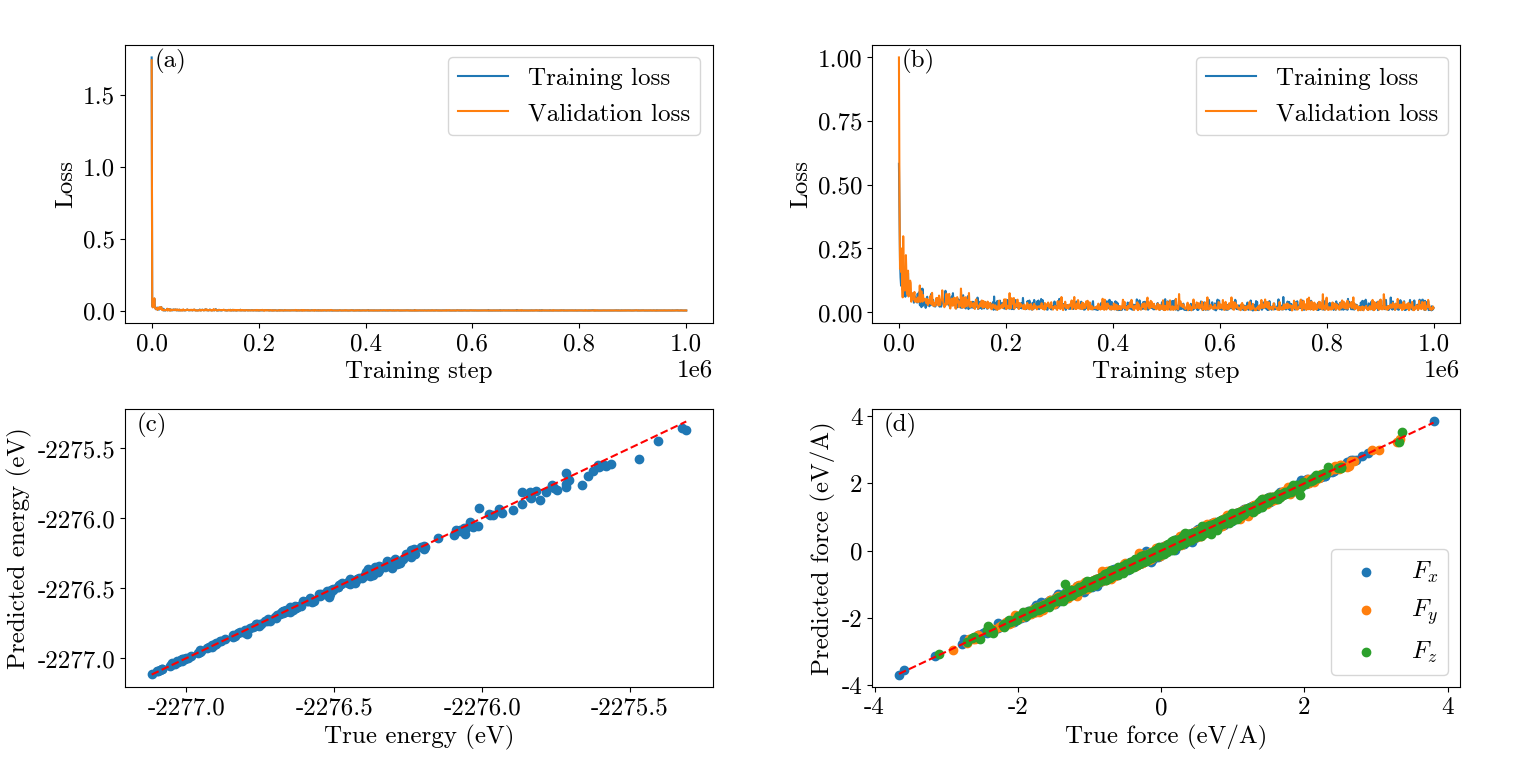}
    \caption{Example results from machine learning on the construction of a machine learning potential model for the C$_{12}$H$_{26}$ molecule by DeePMD-kit. {\bf (a)-(b)} Loss curves from the learning/training process, with training and validation losses both tracked for (a) energy and (b) force loss. {\bf (c)-(d)} Predicted (c) energy and (d) force values by the MLP on frames plotted against the true values from the AIMD simulation.}
    \label{Fig7}
\end{figure*}

The Python-based software package DeePMD-kit\cite{DPMD1,DPMD2} was used to construct machine learning potentials for this work. Specifically, the MLPs are deep learning neural network models known as Deep Potential models \cite{DeeP}. Below, we describe the machine-learning settings and parameters that are consistently applied across all trained MLPs for the molecules examined in this study.

From the AIMD simulations of each molecule (see Sec. \ref{Method-sec}), a total of 1000 frames (consisting of atomic coordinates, energies, and force) were extracted to serve as input data for machine learning. The dataset was divided into 80\% for training and 20\% for validation. During training, each molecular structure is first converted into atomic descriptors, which serve as the input features for the neural network.
The atomic descriptors used are the two-body embedding Deep Potential Smooth Edition (DeepPot-SE) descriptors, which encode the local environment of each atom based on distances to neighboring atoms within a specified cutoff \cite{DPMD1,DPMD2}. 

The extracted atomic descriptors are fed into the fitting neural network to predict the total potential energy, expressed as a sum of atomic contributions, $E=\sum_i E_i$. 
Atomic forces are then obtained as the gradient of the total energy with respect to atomic coordinates. The neural network architecture used is consistent across all models: a three-layer network with 100 neurons per layer, incorporating residual connections (ResNet architecture) between layers. The tanh activation function is applied throughout. Training is performed by minimizing a combined, weighted loss function for both energies and forces, comparing the network predictions to the AIMD reference data. Each network is trained for $10^6$ steps using an exponentially decaying learning rate, which starts at 0.001.

Fig. \ref{Fig7} shows results from the training of one MLP model for dodecane. For both energy and force, loss values were tracked during training to display the loss curves. The trained model was then tested, with predicted energy and force values plotted against 'true' values from the AIMD simulation. These results reflect a ML model that had no fitting issues from the loss curves and performed well in its predictions, which was consistent throughout other models and molecules. However, as discussed in Sec. \ref{Method-sec} about the challenges in this procedure, confirming that the models perform well on data from AIMD does not indicate it is fully prepared to be used in the classical MD timescale, leading to the molecule stability issues.

\renewcommand{\theequation}{B\arabic{equation}}
\renewcommand{\thesection}{B\arabic{section}}
\renewcommand{\thesubsection}{B\arabic{subsection}}
\setcounter{equation}{0}
\setcounter{section}{0} 
\setcounter{subsection}{0} 
\section*{Appendix B: Vibrational spectra analysis of molecules}\label{app:2}

\begin{figure}
    \centering
    \includegraphics[width=\linewidth]{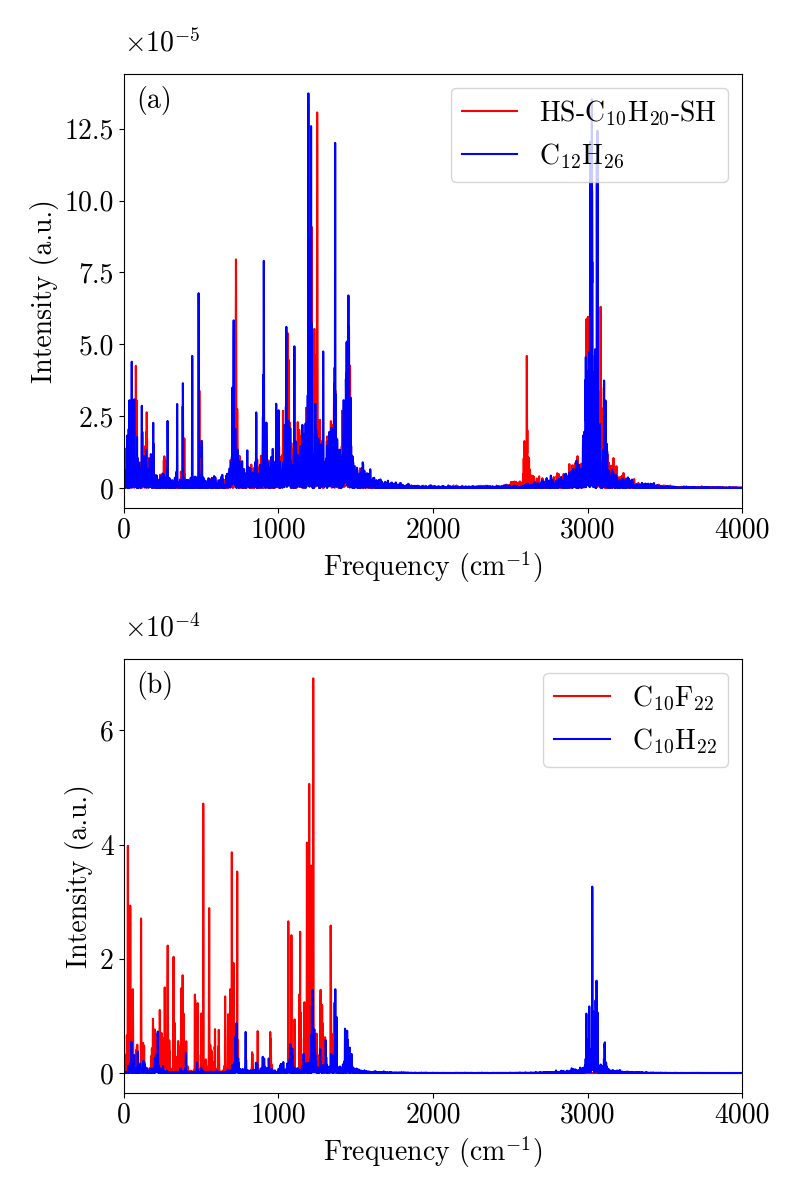}
    \caption{Select simulated vibrational spectra of examined molecules in this work. {\bf (a)} Overlaid spectra of dodecane and decanedithol, corresponding to two characterized molecules in Sec. \ref{End-sec}. {\bf (b)} Overlaid spectra of decane and perfluorodecane, corresponding to two characterized molecules in Sec. \ref{Halo-sec}.}
    \label{Fig8}
\end{figure}

To examine the MLPs and to provide insight into characteristic vibrational modes, Fig. \ref{Fig8} shows simulated vibrational spectra obtained from atomic velocities during NVE MD runs using the trained MLPs. 
Fig. \ref{Fig8}(a) presents the overlaid vibrational spectra of two molecules whose thermal conductance results were discussed in Sec. \ref{End-sec} (different end groups), while Fig. \ref{Fig8}(b) corresponds to the molecules examined in Sec. \ref{Halo-sec} (pure alkane vs fully fluorinated perfluorodecane).

The primary contributing modes to thermal transport are expected to arise from C–C stretching motions. However, these modes are often not mentioned in conventional infrared spectroscopy because they are ubiquitous in organic compounds and span a wide range near the fingerprint region. Indeed, our spectra display numerous peaks in the 800–1500 cm$^{-1}$
region, which may also include contributions from other bonds, depending on the specific molecule. Therefore, it is difficult to draw definitive conclusions about vibrational mismatches affecting thermal transport from these spectra. Instead, these results primarily demonstrate that the MLPs  reproduce distinguishable characteristic vibrational peaks.

In Fig. \ref{Fig8}(a), intense peaks for dodecane (blue) show at $\sim$3000 cm$^{-1}$ (C-H stretch) and $\sim$1300-1400 cm$^{-1}$ (C-H bend). Similar peaks for decanedithiol (red) can be seen, corresponding to the same alkane backbone. A clear S–H stretching mode also shows at $\sim$2600 cm$^{-1}$. For perfluorodecane in Fig. \ref{Fig8}(b), 
high-frequency peaks of C–H stretching are absent, while strong peaks appear around  $\sim$1300 cm$^{-1}$ indicating C-F stretching, as well as other low-frequency modes compared to decane.

\renewcommand{\theequation}{C\arabic{equation}}
\renewcommand{\thesection}{C\arabic{section}}
\renewcommand{\thesubsection}{C\arabic{subsection}}
\setcounter{equation}{0}
\setcounter{section}{0} 
\setcounter{subsection}{0} 
\section*{Appendix C: 
Generalizability of machine learning potentials 
}
\label{app:3}

\begin{figure}
    \centering
    \includegraphics[width=\linewidth]{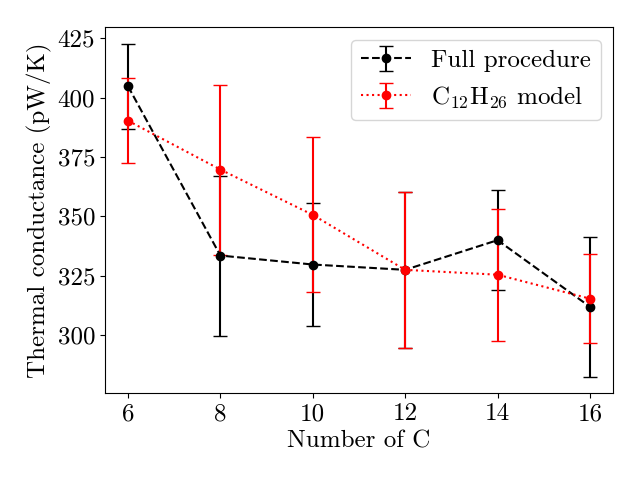}
    \caption{Thermal conductance of pure alkane chains (C$_{N}$H$_{2N+2}$) as a function of number of carbons ($N$). Results obtained through the full MLP procedure (black) are compared to results from applying the trained C$_{12}$H$_{26}$ MLP to other lengths (red). }
    \label{Fig9}
\end{figure}

We tested the generalizability of the machine-learning model by applying a set of MLPs trained solely on AIMD data of C$_{12}$H$_{26}$ to study thermal transport through alkane molecules of different chain lengths.
The resulting $G_{\rm th}$ values are shown in red in Fig. \ref{Fig9}. Compared to results obtained from the full training procedure (black), the
 C$_{12}$H$_{26}$-trained models exhibit good generalizability, with predicted conductances generally falling within the error range of repeated simulations using multiple models. For short chains with $N=6$, the C$_{12}$H$_{26}$ models predict an enhanced conductance, consistent with the full training results.
 For longer chains, the C$_{12}$H$_{26}$ models predicts 
 an approximately linear decrease in conductance from 
 C$_{6}$H$_{14}$ to C$_{12}$H$_{26}$, while the full-training results show early saturation of the conductance. 

While these results are encouraging, more testing is needed to assess generalizability, such as repeating these simulations with models trained on other lengths.


\newpage

\begin{thebibliography}{99}


\bibitem{RevLu08}
J.-S. Wang, J. Wang, and J. T. L\"u,
"Quantum thermal transport in nanostructures,"
\href{https://doi.org/10.1140/epjb/e2008-00195-8}{Eur. Phys. J. B}
{\bf 62}, 381 (2008).

\bibitem{Dhar}
A. Dhar, "Heat transport in low-dimensional systems," \href{https://doi.org/10.1080/00018730802538522}{Adv. Phys.} {\bf 57}, 457 (2008).

\bibitem{Pop10}
E. Pop,
"Energy dissipation and transport in nanoscale devices,"
\href{https://doi.org/10.1007/s12274-010-1019-z}{Nano Res.} {\bf 3}, 147 (2010).


\bibitem{Baowen12}
N. Li, J. Ren, L. Wang, G. Zhang, P. H\"anggi, and B. Li,
"Colloquium: Phononics: Manipulating heat flow with electronic analogs and beyond,"
\href{https://doi.org/10.1103/RevModPhys.84.1045}{Rev. Mod. Phys.} {\bf 84}, 1045 (2012).

\bibitem{Luo13}
T. Luo and G. Chen, 
"Nanoscale Heat Transfer - from Computation to Experiment," 
\href{https://doi.org/10.1039/C2CP43771F}{Phys. Chem. Chem. Phys.} {\bf 15}, 3389 (2013).

\bibitem{Rev14}
D. G. Cahill,  P. V. Braun, G. Chen, D. R. Clarke, S. Fan, K. E. Goodson, P. Keblinski, W. P. King, G. D. Mahan, A. Majumdar, H. J. Maris, S. R. Phillpot, E. Pop, and L. Shi,
"Nanoscale thermal transport. II. 2003–2012,"
\href{https://doi.org/10.1063/1.4832615}{Appl. Phys. Rev.} {\bf 1}, 011305 (2014).

\bibitem{Leitner15}
D. M. Leitner, 
"Quantum Ergodicity and Energy Flow in Molecules,"
\href{https://doi.org/10.1080/00018732.2015.1109817}{Adv. Phys.} {\bf 64}, 445 (2015).

\bibitem{RevA}
D. Segal and  B. K. Agarwalla,
"Vibrational Heat Transport in Molecular Junctions,"
\href{https://doi.org/10.1146/annurev-physchem-040215-112103}{Ann. Rev. Phys. Chem.} {\bf 67}, 185 (2016).

\bibitem{Rubtsov19}
I. V. Rubtsov and A. L. Burin,
"Ballistic and diffusive vibrational energy transport in molecules,"
\href{https://doi.org/10.1063/1.5055670}{J. Chem. Phys.} {\bf 150}, 020901 (2019).

\bibitem{Yoon20}
S. Park, J. Jang, H. Kim, D. I. Park, K. Kim, and H. J. Yoon,
"Thermal conductance in single molecules and self-assembled monolayers: physicochemical insights, progress, and challenges,"
\href{https://doi.org/10.1039/D0TA07095E}{J. Mater. Chem. A} {\bf 8}, 19746 (2020).

\bibitem{HuRev21}
M. Hu and Z. Yang, "Perspective on multi-scale simulation of thermal transport in solids and interfaces," \href{https://doi.org/10.1039/D0CP03372C}{Phys. Chem. Chem. Phys.} {\bf 23}, 1785 (2021).

\bibitem{BaowenR21}
Y. Li, W. Li, T. Han, X. Zheng, J. Li, B. Li, S. Fan, and C.-W. Qiu,
"Transforming heat transfer with thermal metamaterials and devices,"
\href{https://doi.org/10.1038/s41578-021-00283-2}{Nat. Rev. Mater.} {\bf 6}, 488 (2021).

\bibitem{Baowen22}
J. Chen, X. Xu, J. Zhou, and B. Li, "Interfacial thermal resistance: Past, present, and future," \href{https://doi.org/10.1103/RevModPhys.94.025002}{Rev. Mod. Phys.} {\bf 94}, 025002 (2022).

\bibitem{RevG}
B. Gotsmann, A. Gemma, and D. Segal,
"Quantum phonon transport through channels and molecules—A Perspective,"
\href{https://doi.org/10.1063/5.0088460}{App. Phys. Lett.} {\bf 120}, 160503 (2022).

\bibitem{Rev23}
S. Chatterjee, Paras, H. Hu, and M. Chakraborty, "A Review of Nano and Microscale Heat Transfer: An Experimental and Molecular Dynamics Perspective," \href{A Review of Nano and Microscale Heat Transfer: An Experimental and Molecular Dynamics Perspective}{Processes} {\bf 11}, 2769 (2023).

\bibitem{Rev25}
J. Kim, Y. Liu, T. Luo, and Z. Tian, "Molecular Dynamics Simulations in Nanoscale Heat Transfer: A Mini Review," \href{https://doi.org/10.1115/1.4067341}{J. Heat Mass Transfer.} {\bf 147}, 030801 (2025).



\bibitem{Wang06}
R. Y. Wang, R. A. Segalman, and A. Majumdar,
"Room temperature thermal conductance of alkanedithiol self-assembled monolayers,"
\href{https://doi.org/10.1063/1.2358856}{Appl. Phys. Lett.} {\bf 89}, 173113 (2006).

\bibitem{Dlott07}
Z. Wang, J. A. Carter, A. Lagutchev, Y. K. Koh, N.-H. Seong, D. G. Cahill, and D. D. Dlott,
"Ultrafast Flash Thermal Conductance of Molecular Chains,"
\href{https://doi.org/10.1126/science.1145220}{Science} {\bf 317}, 787 (2007).

\bibitem{Cahill12}
M. D. Losego, M. E. Grady, N. R. Sottos, D. G. Cahill, and P. V. Braun,
"Effects of chemical bonding on heat transport across interfaces,"
\href{https://doi.org/10.1038/nmat3303}{Nat. Mater.} {\bf 11}, 502 (2012).

\bibitem{GotsmannExp14}
T. Meier, F. Menges, P. Nirmalraj, H. H\"olscher, H. Riel, and B. Gotsmann,
"Length-Dependent Thermal Transport along Molecular Chains,"
\href{https://doi.org/10.1103/PhysRevLett.113.060801}{Phys. Rev. Lett.} {\bf 113}, 060801 (2014).

\bibitem{Shub15}
S. Majumdar, J. A. Sierra-Suarez, S. N. Schiffres, W.-L. Ong, C. F. Higgs III, A. J. H. McGaughey, and J. A. Malen,
"Vibrational Mismatch of Metal Leads Controls Thermal Conductance of Self-Assembled Monolayer Junctions,"
\href{https://doi.org/10.1021/nl504844d}{Nano Lett.} {\bf 15}, 2985 (2015).

\bibitem{Shub17}
S. Majumdar, J. A. Malen, and A. J. H. McGaughey,
"Cooperative Molecular Behavior Enhances the Thermal Conductance of Binary Self-Assembled Monolayer Junctions,"
\href{https://doi.org/10.1021/acs.nanolett.6b03894}{Nano Lett.} {\bf 17}, 220 (2017).

\bibitem{Rubtsov21}
S. U. Nawagamuwage L. N. Qasim, X. Zhou, T. X. Leong, I. V. Parshin, J. Jayawickramarajah A. L. Burin, and I. V. Rubtsov,
"Competition of Several Energy-Transport Initiation Mechanisms Defines the Ballistic Transport Speed,"
\href{https://doi.org/10.1021/acs.jpcb.1c03986}{J. Phys. Chem. B} {\bf 125}, 7546 (2021).


\bibitem{Rubtsov24}
S. U. Nawagamuwage, E. S. Williams, M. M. Islam, I. V. Parshin, A. L. Burin, N. Busschaert, and I. V. Rubtsov,
"Ballistic energy transport via long alkyl chains: A new initiation mechanism,"
\href{https://doi.org/10.1021/acs.jpcb.4c03386}{J. Phys. Chem. B} {\bf 128}, 8788 (2024).
\bibitem{CuiExp19}
L. Cui, S. Hur, Z. A. Akbar, J. C. Kl\"ockner, W. Jeong, F. Pauly, S.-Y. Jang, P. Reddy, and E. Meyhofer,
"Thermal conductance of single-molecule junctions,"
\href{https://doi.org/10.1038/s41586-019-1420-z}{Nature} {\bf 572}, 628 (2019).

\bibitem{GotsmannExp19}
N. Mosso, H. Sadeghi, A. Gemma, S. Sangtarash, U. Drechsler, C. Lambert, and B. Gotsmann,
"Thermal Transport through Single-Molecule Junctions,"
\href{https://doi.org/10.1021/acs.nanolett.9b02089}{Nano Lett.} {\bf 19}, 7614 (2019).

\bibitem{GotsmannExp23}
A. Gemma, F. Tabatabaei, U. Drechsler, A. Zulji, H. Dekkiche, N. Mosso, T. Niehaus, M. R. Bryce, S. Merabia, and B. Gotsmann, "Full thermoelectric characterization of a single molecule," \href{https://doi.org/10.1038/s41467-023-39368-7}{Nat. Commun.} {\bf 14}, 3868 (2023).

\bibitem{CuiExp25}
S. C. Yelishala, Y. Zhu, P. M. Martinez, H. Chen, M. Habibi, G. Prampolini, J. C. Cuevas, W. Zhang, J. G. Vilhena, and L. Cui, "Phonon interference in single-molecule junctions," \href{https://doi.org/10.1038/s41563-025-02195-w}{Nat. Mater.} {\bf 24}, 1258 (2025).






\bibitem{Dvira2003}
D. Segal, A. Nitzan, and P. H\"anggi,
"Thermal conductance through molecular wires,"
\href{https://doi.org/10.1063/1.1603211}{J. Chem. Phys.} {\bf 119}, 6840 (2003).

\bibitem{SegalQME}
D. Segal,
"Heat flow in nonlinear molecular junctions: Master equation analysis ",
\href{https://link.aps.org/doi/10.1103/PhysRevB.73.205415}
{Phys. Rev. B}, {\bf 73}, 205415 (2006). 

\bibitem{SegalSB}
D. Segal,
"Heat transfer in the spin-boson model: A comparative study in the incoherent tunneling regime,"
\href{https://doi.org/10.1103/PhysRevE.90.012148}{
Phys. Rev. E} {\bf 90}, 012148 (2014).

\bibitem{Dhar06}
A. Dhar and D. Roy, 
"Heat transport in harmonic lattices,"
\href{https://doi.org/10.1007/s10955-006-9235-3}{J. Stat. Phys.} {\bf 125}, 801 (2006).

\bibitem{Malay11}
M. Bandyopadhyay, and D. Segal,
"Quantum heat transfer in harmonic chains with self-consistent reservoirs: Exact numerical simulations,"
\href{https://doi.org/10.1103/PhysRevE.84.011151}{Phys. Rev. E} {\bf 84}, 011151 (2011).

\bibitem{Tulkki13}
K. S\"a\"askilahti, J. Oksanen, and J. Tulkki,
"Thermal balance and quantum heat transport in nanostructures thermalized by local Langevin heat baths,"
\href{https://journals.aps.org/pre/abstract/10.1103/PhysRevE.88.012128}{
Phys. Rev. } E {\bf 88}, 012128 (2013).
%

\bibitem{Tulkki15}
K. S\"a\"askilahti, J. Oksanen, S. Volz, and J. Tulkki,
"Nonequilibrium phonon mean free paths in anharmonic chains,"
\href{https://doi.org/10.1103/PhysRevB.92.245411}
{Phys. Rev. B} {\bf 92}, 245411 (2015).

\bibitem{Roya19}
R. Moghaddasi Fereidani and D. Segal,
"Phononic heat transport in molecular junctions: Quantum effects and vibrational mismatch,"
\href{https://doi.org/10.1063/1.5075620}{J. Chem. Phys.} {\bf 150}, 024105 (2019).

\bibitem{Naim}
N. Kalantar, B. K. Agarwalla, and D. Segal, "On the definitions and simulations of vibrational heat transport in nanojunctions," \href{https://doi.org/10.1063/5.0027414}{J. Chem. Phys.} {\bf 153}, 174101 (2020).


\bibitem{Maly21}
J. Behera and M. Bandyopadhyay,
"Environment-dependent vibrational heat transport in molecular junctions: Rectification, quantum effects, vibrational mismatch,"
Phys. Rev. E {\bf 104}, 014148 (2021).

\bibitem{Markussen13}
T. Markussen, "Phonon interference effects in molecular junctions," \href{https://doi.org/10.1063/1.4849178}{J. Chem. Phys.} {\bf 139}, 244101 (2013).

\bibitem{Pauly16}
J. C. Kl\"ockner, M. B\"urkle, J. C. Cuevas, and F. Pauly,
"Length dependence of the thermal conductance of alkane-based single-molecule junctions: An ab initio study,"
\href{https://doi.org/10.1103/PhysRevB.94.205425}{Phys. Rev. B} {\bf 94}, 205425 (2016).


\bibitem{ChristmasTree}
M. Famili, I. Grace, H. Sadeghi, and C. J. Lambert,
"Suppression of Phonon Transport in Molecular Christmas Trees,"
\href{https://doi.org/10.1002/cphc.201700147}{ChemPhysChem} {\bf 18}, 1234 (2017).


\bibitem{Teflon}
M. Buerkle and Y. Asai,
"Thermal conductance of Teflon and Polyethylene: Insight from an atomistic, single-molecule level,"
\href{https://doi.org/10.1038/srep41898}{Sci. Rep.} {\bf 7}, 41898 (2017).


\bibitem{Pauly17}
J. C. Kl\"ockner, J. C. Cuevas, and F. Pauly,
"Tuning the thermal conductance of molecular junctions with interference effects," \href{https://doi.org/10.1103/PhysRevB.96.245419}{Phys. Rev. B} {\bf 96}, 245419 (2017).


\bibitem{Pauly18}
J. C. Kl\"ockner, J. C. Cuevas, and F. Pauly,
"Transmission eigenchannels for coherent phonon transport,"
\href{https://doi.org/10.1103/PhysRevB.97.155432}{Phys. Rev. B} {\bf 97}, 155432 (2018).


\bibitem{Pauly23}
J. C. Klöckner and F. Pauly, "Variability of the thermal
conductance of gold-alkane-gold single-molecule junctions
studied using ab-initio and molecular dynamics approaches," \href{https://arxiv.org/abs/1910.02443}{arXiv:1910.02443}.


\bibitem{Hatef25}
L. Zheng, E. N. Farahani, A. H. S. Daaoub, S. Sangtarash, and H. Sadeghi,
"Rules of Connectivity-Dependent Phonon Interference in Molecular Junctions," \href{https://doi.org/10.1021/acs.nanolett.5c00225}{Nano. Lett.} {\bf 25}, 6524 (2025).

\bibitem{Heavyatom}
W. Bro-J\o rgensen, A. J. Bay-Smidt, D. Donadio, and G. C. Solomon,
"Heavy Solution for Molecular Thermal Management: Phonon Transport Suppression with Heavy Atoms,"
\href{https://doi.org/10.1021/acsphyschemau.4c00084}{ACS Phys. Chem. Au} {\bf 5}, 162 (2025).

\bibitem{Fabian25}
M. Blaschke and F. Pauly,
"Revealing molecule‑internal mechanisms that control phonon heat transport through single‑molecule junctions by a genetic algorithm,"
\href{https://doi.org/10.1021/acsnano.5c03690}{ACS Nano} {\bf 19}, 32093 (2025).

\bibitem{Pawel11}
K. Sasikumar and P. Keblinski,
"Effect of chain conformation in the phonon
transport across a Si-polyethylene single-
molecule covalent junction,"
\href{https://doi.org/10.1063/1.3592296}{J. Appl. Phys.} {\bf 109}, 114307 (2011). 



\bibitem{NitzanD20}
M. Dinpajooh and A. Nitzan, "Heat conduction in polymer chains with controlled end-to-end distance," \href{https://doi.org/10.1063/5.0023085}{J. Chem. Phys.} {\bf 153}, 164903 (2020).

\bibitem{Nitzan20}
I. Sharony, R. Chen, and A. Nitzan,
"Stochastic simulation of nonequilibrium heat conduction in extended molecular junctions,"
\href{https://doi.org/10.1063/5.0022423}{J. Chem. Phys.} {\bf 153}, 144113 (2020).



\bibitem{Lu2021}
G. Li, B.-Z. Hu, N. Yang, and J.-T. L\"u,
"Temperature-dependent thermal transport of single molecular junctions from semiclassical Langevin molecular dynamics,"
\href{https://doi.org/10.1103/PhysRevB.104.245413}{Phys. Rev. B} {\bf 104}, 245413 (2021).


\bibitem{Nitzan22}
M. Dinpajooh and A. Nitzan,
"Heat conduction in polymer chains: Effect of substrate on the thermal conductance,"
\href{https://doi.org/10.1063/5.0087163}{J. Chem. Phys.} {\bf 156}, 144901 (2022).


\bibitem{Benzene}
B. A. Mart\'inez-Torres, F. Salazar, and M. Romero-Bastida,
"Simulation of nonequilibrium heat conduction in benzene single-molecule junctions,"
\href{https://iopscience.iop.org/article/10.1088/1361-648X/ae1c0e/meta}{J. Phys.: Condens. Matter} {\bf 37}, 465304 (2025).

\bibitem{JW-HeatMD}
J. J. Wang, J. Gong, A. J. H. McGaughey, and D. Segal, "Simulations of heat transport in single-molecule junctions: Investigations of the thermal diode effect," \href{https://doi.org/10.1063/5.0125714}{J. Chem. Phys.} {\bf 157}, 174105 (2022).

\bibitem{JW-HeatFluc}
J. J. Wang, M. Gerry, and D. Segal, "Challenges in molecular dynamics simulations of heat exchange statistics," \href{https://doi.org/10.1063/5.0187357}{J. Chem. Phys.} {\bf 160}, 074111 (2024).

\bibitem{JW-Fullerene}
J. Li, J. J. Wang, and D. Segal, "Thermal transport in fullerene-based molecular junctions: molecular dynamics simulations," \href{https://iopscience.iop.org/article/10.1088/1361-648X/ad459b}{J. Phys.: Condens. Matter} {\bf 36}, 325901 (2024).




\bibitem{ANI}
J. S. Smith, O. Isayev, and A. E. Roitberg,
"ANI-1: An extensible neural network potential with DFT accuracy at force-field computational cost,"
\href{https://doi.org/10.1039/C6SC05720A}{Chem. Sci.} {\bf 8}, 3192 (2017).

\bibitem{DeeP}
L. Zhang, J. Han, H. Wang, R. Car, and W. E, "Deep Potential Molecular Dynamics: A Scalable Model with the Accuracy of Quantum Mechanics," \href{https://doi.org/10.1103/PhysRevLett.120.143001}{Phys. Rev. Lett.} {\bf 120}, 143001 (2018).


\bibitem{MLP19}
H. Chan, B. Narayanan, M. J. Cherukara, F. G. Sen, K. Sasikumar, S. K. Gray, M. K. Y. Chan, and S. K. R. S. Sankaranarayanan,
"Machine Learning Classical Interatomic Potentials for Molecular Dynamics from First-Principles Training Data,"
\href{https://doi.org/10.1021/acs.jpcc.8b09917}{J. Phys. Chem. C} {\bf 123}, 6941 (2019).

\bibitem{MLthermal20}
Y.-B. Liu, J.-Y. Yang, G.-M. Xin, L.-H. Liu, G. Cs\'anyi, and B.-Y. Cao,
"Machine learning interatomic potential developed for molecular simulations on thermal properties of $\beta$-Ga$_2$O$_3$,"
\href{https://doi.org/10.1063/5.0027643}{J. Chem. Phys.} {\bf 153}, 144501 (2020).


\bibitem{MLRev21}
O. T. Unke, S. Chmiela, H. E. Sauceda, M. Gastegger, I. Poltavsky, K. T. Sch\"utt, A. Tkatchenko, and K. R. M\"uller,
Machine Learning Force Fields
Chemical Reviews {\bf 121}, 10142-10186 (2021).

\bibitem{MLrev1}
X. Qian and R. Yang, "Machine learning for predicting thermal transport properties of solids," \href{https://doi.org/10.1016/j.mser.2021.100642}{Mater. Sci. Eng. R Rep.} {\bf 146}, 100642 (2021).


\bibitem{Lu21}
G. Li, B.-Z. Hu, W.-H. Mao, N. Yang,  J.-T. L\"u, 
"Order of magnitude reduction in Joule heating of single molecular junctions between graphene electrodes," 
\href{https://doi.org/10.1063/5.0118952}{
J. Chem. Phys.} {\bf 157}, 174303 (2022).
%





\bibitem{Siesta1}
J. M. Soler, E. Artacho, J. D. Gale, A. Garc\'{i}a, J. Junquera, P. Ordej\'{o}n, and D. S\'{a}nchez-Portal, "The SIESTA method for \textit{ab initio} order-\textit{N} materials simulation," \href{https://iopscience.iop.org/article/10.1088/0953-8984/14/11/302}{J. Phys.: Condens. Matter} {\bf 14}, 2745 (2002).

\bibitem{Siesta2}
A. Garcia, N. Papior, A. Akhtar, E. Artacho, V. Blum, E. Bosoni, P. Brandimarte, M. Brandbyge, J. I. Cerd\'{a}, F. Corsetti, R. Cuadrado, V. Dikan, J. Ferrer, J. Gale, P. Garc\'{i}a-Fern\'{a}ndez, V. M. Garc\'{i}a-Su\'{a}rez, S. Garc\'{i}a, G. Huhs, S. Illera, R. Koryt\'{a}r, P. Koval, I. Lebedeva, L. Lin, P. L\'{o}pez-Tarifa, S. G. Mayo, S. Mohr, P. Ordej\'{o}n, A. Postnikov, Y. Pouillon, M. Pruneda, R. Robles, D. S\'{a}nchez-Portal, J. M. Soler, R. Ullah, V. W. Yu, and J. Junquera, "SIESTA: Recent developments and applications," \href{https://doi.org/10.1063/5.0005077}{J. Chem. Phys.} {\bf 152}, 204108 (2020). 

\bibitem{DPMD1}
H. Wang, L. Zhang, J. Han, and W. E, "DeePMD-kit: A deep learning package for many-body potential energy representation and molecular dynamics," \href{https://doi.org/10.1016/j.cpc.2018.03.016}{Comput. Phys. Commun.} {\bf 228}, 178 (2018).

\bibitem{DPMD2}
J. Zeng, D. Zhang, D. Lu, P. Mo, Z. Li, Y. Chen, M. Rynik, L. Huang, Z. Li, S. Shi, Y. Wang, H. Ye, P. Tuo, J. Yang, Y. Ding, Y. Li, D. Tisi, Q. Zeng, H. Bao, Y. Xia, J. Huang, K. Muraoka, Y. Wang, J. Chang, F. Yuan, S. L. Bore, C. Cai, Y. Lin, B. Wang, J. Xu, J.-X. Zhu, C. Luo, Y. Zhang, R. E. A. Goodall, W. Liang, A. K. Singh, S. Yao, J. Zhang, R. Wentzcovitch, J. Han, J. Liu, W. Jia, D. M. York, W. E, R. Car, L. Zhang, and H. Wang, "DeePMD-kit v2: A software package for deep potential models," \href{https://doi.org/10.1063/5.0155600}{J. Chem. Phys.} {\bf 159}, 054801 (2023).

\bibitem{LAMMPS}
A. P. Thompson, H. M. Aktulga, R. Berger, D. S. Bolintineanu, W. M. Brown, P. S. Crozier, P. J. in 't Veld, A. Kohlmeyer, S. G. Moore, T. D. Nguyen, R. Shan, M. J. Stevens, J. Tranchida, C. Trott, and S. J. Plimpton,
"LAMMPS - a flexible simulation tool for particle-based materials modeling at the atomic, meso, and continuum scales,"
\href{https://doi.org/10.1016/j.cpc.2021.108171}{Comp. Phys. Comm.} {\bf 271}, 10817 (2022).
\bibitem{OVITO}
A. Stukowski, "Visualization and analysis of atomistic simulation data with OVITO-the Open Visualization Tool," \href{https://iopscience.iop.org/article/10.1088/0965-0393/18/1/015012}{Modelling Simul. Mater. Sci. Eng.} {\bf 18}, 015012 (2010).





\bibitem{electronG}
 Y. F. Wang, J. Kröger, R. Berndt, H. Vázquez, M. Brandbyge, and M. Paulsson,
 "Atomic-Scale Control of Electron Transport through Single Molecules,"
 \href{https://doi.org/10.1103/PhysRevLett.104.176802}{Phys. Rev. Lett.} {\bf 104}, 176802 (2010). 

 \bibitem{Leitner19}
K. M. Reid, H. D. Pandey, D. M. Leitner,
"Elastic and Inelastic Contributions to Thermal Transport between Chemical Groups and Thermal Rectification in Molecules,"
\href{https://pubs.acs.org/doi/full/10.1021/acs.jpcc.8b11640}
{J. Phys. Chem. C}  {\bf 123}, 6256-6264 (2019). 

\bibitem{LeitnerB}
D. M. Leitner, H. D. Pandey, K. M. Reid. 
"Energy Transport across Interfaces in Biomolecular Systems," 
\href{https://doi.org/10.1021/acs.jpcb.9b07086}{
The Journal of Physical Chemistry B} {\bf 123}, 9507-9524 (2019).






 















\end{thebibliography}
\end{document}